\documentclass[amstex]{article}
\usepackage{graphicx}
\usepackage{dcolumn}
\usepackage{amsmath}
\usepackage{amssymb}
\usepackage{amsfonts}
\usepackage{amscd}
\begin{document}
\newtheorem{thm}{Theorem}[section]
\newtheorem{lem}[thm]{Lemma}
\newtheorem{prop}[thm]{Proposition}
\newtheorem{cor}[thm]{Corollary}
\newtheorem{assum}{Assumption}[section]
\newtheorem{rem}[thm]{Remark}
\newtheorem{defn}[thm]{Definition}
\newtheorem{rthm}{Theorem}
\newcommand{\lv}{\left \vert}
\newcommand{\rv}{\right \vert}
\newcommand{\la}{\left \langle}
\newcommand{\ra}{\right \rangle}
\newcommand{\ket}[1]{\lv #1 \ra}
\newcommand{\bra}[1]{\la #1 \rv}

\begin{center}
\bigskip\bigskip
{\huge\bf
Local distinguishability of\\}
\bigskip
{\huge\bf 
quantum states \\}
\bigskip
{\huge\bf in infinite dimensional
systems\\}
\bigskip\bigskip
\bigskip\bigskip
{\Large Yoshiko Ogata}
\\
\bigskip
{\it Department of Mathematical Sciences,
The University of Tokyo, 
3-8-1 Komaba, Meguro, Tokyo, 153-8914 Japan}
\end{center}
\noindent
We investigate local distinguishability of quantum states by
use of the convex analysis about
{\it joint numerical range} of operators on a Hilbert space.
We show that any two orthogonal pure states are distinguishable
by local operations and classical communications,
even for infinite dimensional systems.
An estimate of the local 
discrimination probability is also given for some family of 
more than two pure states.
\noindent
\bigskip
\noindent
\section{Introduction}
\quad
Local operations and classical communications (LOCC)
are basic operations in quantum information theory.
Many interesting studies have arisen from the question, what we 
can$\backslash$cannot do using only LOCC.
The question is highly non-trivial and difficult to solve due to the lack of simple characterization of LOCC.
The necessary and sufficient condition of the deterministic convertibility of one pure state to the other was 
derived by Nielsen, for general bipartite systems, in \cite{Nielsen}.
Furthermore, in \cite{Vidal},
Vidal obtained the optimal probability to convert one pure state
to the other, non-deterministically.
However, when we start to think of simultaneous convertibility
of more than one states, the problem becomes furthermore difficult, because of
the fact that Lo-Popescu Theorem \cite{lop} is not applicable there.

The local distinguishability problem is one of 
these questions.
The problem is as follows:
We investigate a combined quantum system consisting of two parts 
$A$ and $B$ held by separated observers (Alice and Bob).
We denote the associated Hilbert space by ${\cal H}_A\otimes{\cal H}_B$,
where ${\cal H}_A$, ${\cal H}_B$ are separable 
(i.e., possibly infinite dimensional) Hilbert spaces
that represent the system of Alice and Bob,
respectively.
Let $\psi_1,\cdots,\psi_M$ be orthonormal vectors
in ${\cal H}_A\otimes{\cal H}_B$, which represent
$M$ pure states.
Suppose that the system is in a state $\psi$,
which is prepared to be 
one of $\psi_1,\cdots,\psi_M$.
Alice and Bob know that $\psi$ is one of 
$\psi_1,\cdots,\psi_M$,
but they don't know which of them it is.
The problem is if Alice and Bob can find out which one it is,
when only LOCC is allowed.

In \cite{WH}, Walgate et.al. proved that any two orthogonal pure states in finite dimensional systems are distinguishable.
Unfortunately, because of the nature of 
their proof, this important result has been restricted 
to finite dimensional systems so far. 
As it is indispensable to consider infinite 
dimensional systems in the real world, 
the analogous result in infinite dimensional system is desirable.
In this paper, we prove the infinite version:
\newtheorem{wthm}{Theorem}[section]
\renewcommand{\thewthm}{}
\begin{wthm}
Any two orthogonal pure states are distinguishable
by LOCC, even for infinite dimensional systems.
\end{wthm}

In spite of these simple results 
for two pure states, it is known that 
more than two pure states are not always distinguishable by LOCC.
It was proved that three Bell states can not be distinguished
with certainty by LOCC and four Bell states can not, even
probabilistically \cite{Ghosh}.
A set of non-entangled pure states 
that are not locally distinguishable was introduced in \cite{ben}.
The probability of the discrimination for the worst case
was estimated in \cite{na}.
In this paper, we give an estimate of discrimination probability for
some family of more than two pure states. This result also holds 
for infinite dimensional systems.\\

In order to investigate distinguishability, we look for a
suitable decomposition of the states.
Let us decompose the vectors 
$\psi_1,\cdots,\psi_M$
with respect to an orthonornal basis $\{e_k\}$ of 
${\cal H}_B$: 
\begin{align}
\psi_l=\sum_k\xi_k^l\otimes e_k ,\quad l=1,\cdots, M.
\label{decom}
\end{align}
(Here and below,
if the dimension of ${\cal H}_B$ is finite $n$,
$\sum_i \varphi'_i\otimes f_i'$ stands for 
the sum $\sum_{i=1}^n\varphi'_i\otimes f_i'$,
while if ${\cal H}_B$ is infinite dimensional,
it stands for the limit
$\lim_{n\to\infty}\sum_{i=1}^n \varphi_i'\otimes f_i'$, when
the limit converges 
in the norm topology of ${\cal H}_A\otimes{\cal H}_B$.)
Suppose that the vectors $\{\xi_k^l\}$ satisfy the
orthogonal conditions for each $k$:
\begin{align}
\langle \xi_k^l\ket{\xi_k^m}=0\quad\forall{l\neq m},
\quad\forall k.
\label{ortho}
\end{align}
This orthogonality condition does not hold in general,
but if this condition holds, 
Alice and Bob can distinguish these states by the following LOCC:
First Bob performs a projective measurement 
$\{\ket{e_k}\bra{e_k}\}$ on his side. 
Then he tells the result $k$ of his measurement
to Alice by a classical communication.
For each $k$, let $S_k$ be a set of $1\le l\le M$ such that $\xi_k^l\neq0$.
According to the information from Bob, Alice performs a projective measurement 
given by projections
$\{\ket{\hat\xi_k^l}\bra{\hat\xi_k^l}\}_{l\in{S_k}}$ and
$1-\sum_{l\in{S_k}}\ket{\hat\xi_k^l}\bra{\hat\xi_k^l}$.
Here, a vector $\hat\xi_k^l\in{\cal H}_A$ 
is the normalization of the vector
$\xi_k^l\in{\cal H}_A$.
As $\{\xi_k^l\}_{l\in S_k}$ are mutually orthogonal
for each $k$,
the projections are orthogonal.
Because the initial state $\psi$ was prepared to be
one of $\psi_1,\cdots,\psi_M$,
Alice obtains one of $\hat\xi_k^l$, $l\in{S_k}$.
When Bob obtains $e_k$ and Alice obtains $\hat{\xi_k^l}$,
they can say the original state $\psi$ was $\psi_l$,
because if $\psi=\psi_m$ for $m\neq l$,
the probability that they obtain $e_k$ and $\hat{\xi_k^l}$
is $0$.
Hence a deterministic local discrimination is possible when the decomposition
(\ref{decom}) with the orthogonality condition (\ref{ortho})
is given.

Next let us consider probabilistic discriminations.
Suppose that $\psi_1,\cdots,\psi_M$ are decomposed into the form
(\ref{decom}),
but now the orthogonal condition holds only partially,
i.e., just for $k$ larger than some $N_p$:
\begin{align}
\langle \xi_k^l\ket{\xi_k^m}=0\quad 
\forall{l\neq m}\quad\quad 
\forall k> N_p.
\label{prob}
\end{align}
In this case, $\psi_1,\cdots,\psi_M$ can be 
distinguished by conclusive LOCC protocol,
 probabilistically.
Let ${P}_{d}$ be the largest probability that can be attained.
The conclusive protocol below gives the lower bound of 
${P}_d$:
\begin{align}
P_d\ge 1-\underset{1\le l\le M}{\max}
\sum_{k=1}^{N_p}\Vert\xi_k^l\Vert^2.
\label{pb}
\end{align}First Bob performs the projective measurement $\{\ket{e_k}\bra{e_k}\}$ again.
If he gets the result $k> N_p$, he tells the result to Alice.
Then Alice performs the projective measurement given by projections
$\{\ket{\hat\xi_k^l}\bra{\hat\xi_k^l}\}_{l\in{S_k}}$
and
$1-\sum_{l\in S_k}\ket{\hat\xi_k^l}\bra{\hat\xi_k^l}$,
and obtains one of $\hat\xi_k^l$.
If she gets $\hat\xi_k^l$, then
they can conclude $\psi=\psi_l$, as before.
In this way, they can distinguish $\psi_1,\cdots,\psi_M$ 
if the result of Bob's measurement is $k> N_p$.
On the other hand, if Bob obtains $k\le N_p$, 
we regard it as an error.
When $\psi=\psi_l$, the probability the error occurs is 
$\sum_{k=1}^{N_p}\Vert\xi_k^l\Vert^2$.
Hence $\psi_1,\cdots,\psi_M$ can be distinguished by
LOCC with probability $P_d$, lower bounded as in (\ref{pb}).

The problem is if there is a decomposition (\ref{decom})
of $\psi_1,\cdots,\psi_M$, satisfying the orthogonality condition
(\ref{ortho}) or (\ref{prob}).
In order to deal with this problem, 
we will introduce a real vector space of trace class self-adjoint operators 
on the Hilbert space ${\cal H}_B$, determined by the states $\psi_1,\cdots,\psi_M$.
We will denote the vector space by $\cal K$.
Let $N$ be the dimension of $\cal K$ and $(A_1,\cdots, A_N)$ a basis of $\cal K$.
For every orthogonal projection $P$ on ${\cal H}_B$, 
we investigate the subset of ${\mathbb R}^N$ given by
\begin{align*}
\left\{\left(\left\langle z,A_1 z\right\rangle,
\cdots,\left\langle z,A_N z\right\rangle\right)\; :
\; z\in P{\cal H}_B,\; \Vert z\Vert=1\right\}\subset{\mathbb R}^N.
\end{align*}
This set is the {\it joint numerical range} of
operators $(A_1,\cdots,A_N)$, restricted on the sub-Hilbert space
$P{\cal H}_B$.
We will show that the convexity of these sets implies 
the existence of the decomposition (\ref{decom}) with 
the orthogonality condition (\ref{ortho}),
hence the local
distinguishability of
the states $\psi_1,\cdots,\psi_M$.
One of the advantage of this method is that we can consider infinite
systems, easily.

In this paper, we prove the infinite version of \cite{WH}:
by the convex analysis on joint numerical ranges,
we show that any two pure orthogonal states can be decomposed 
as in (\ref{decom}), with the orthogonality condition (\ref{ortho}).
We also apply our method to investigate the distinguishability of more than two pure states.
We show that if the dimension of $\cal K$ is $3$, 
the condition (\ref{prob})
holds for $N_P=2$, hence the states are distinguishable
probabilistically. (Theorem \ref{snd})

The remainder of the paper is organized in the following way:
In Section \ref{rep}, we introduce a representation of a vector in 
${\cal H}_A\otimes{\cal H}_B$ as an operator 
from ${\cal H}_B$ to ${\cal H}_A$.
And from them,
we define the real vector space $\cal K$.
Then we represent our main results in terms of the vector space $\cal K$.
In Section \ref{proof},
by use of convex analysis on joint numerical ranges, 
we show the distinguishability of states.

\section{The distinguishability of states}\label{rep}
In this section, we introduce a representation of pure states on 
${\cal H}_A\otimes{\cal H}_B$
as operators from ${\cal H}_B$ to ${\cal H}_A$,
and describe our main results in terms of the operator representation.
In finite dimensional systems, the operator representation corresponds to the well known matrix representation of states, by use of a maximal entangled state. 
(See for example \cite{mat}).

Let ${\cal H}_A$, ${\cal H}_B$ be separable 
(possibly infinite dimensional) Hilbert spaces.
Let us fix some orthonormal basis $\{f_i\}$ of 
${\cal H}_B$.
A vector $\psi$ in ${\cal H}_A\otimes{\cal H}_B$ 
can be decomposed as
\begin{align*}
\psi=\sum_i\varphi_i\otimes f_i,
\end{align*}
in general.
Here, the limit $\lim_{n\to\infty}\sum_{i=1}^n\varphi_i\otimes f_i$ converges in the norm topology 
of ${\cal H}_A\otimes{\cal H}_B$
for infinite dimensional case.
The vectors $\varphi_i$ in ${\cal H}_A$ satisfy 
\begin{align}
\sum_i\Vert\varphi_i\Vert^2=\Vert\psi\Vert^2.
\label{psum}
\end{align}
Now we define a bounded linear operator $X$ from ${\cal H}_B$
to ${\cal H}_A$ by
\begin{align}
X\eta\equiv 
\sum_i \langle f_i\vert \eta\rangle\cdot\varphi_i,\quad
\forall\eta\in {\cal H}_B.
\label{Xdef}
\end{align}
From (\ref{psum}), the sum in (\ref{Xdef}) 
absolutely converges
in norm of ${\cal H}_B$,
and we obtain $\Vert X\Vert\le \Vert\psi\Vert$.
Then the vector $\psi$ is represented as
\begin{align*}
\psi=\sum_{i}\varphi_i\otimes f_i
=\sum_i(Xf_i)\otimes f_i.
\end{align*}
The bounded operator $X^*X$ on ${\cal H}_B$ satisfies
\begin{align}
Tr X^*X
=\sum_i\Vert\varphi_i\Vert^2
=\Vert \psi\Vert^2 <\infty,
\label{trace}
\end{align}
i.e., $X^*X$ is a trace class operator on ${\cal H}_B$.
By operating $1\otimes\ket{f_i}\bra{f_i}$ on $\psi$,
we see that $X$ is the unique operator such that
$\psi=\sum_{i}Xf_i\otimes f_i$. 
On the other hand, for any bounded linear operator $X$
from ${\cal H}_B$ to ${\cal H}_A$ satisfying 
$TrX^*X<\infty$, 
there exists a unique vector 
$\sum_{i}X f_i\otimes f_i$,
(i.e., there exists the limit $\lim_{n\to\infty}\sum_{i=1}^nX f_i\otimes f_i$
for infinite dimensional case,
in the norm of ${\cal H}_A\otimes{\cal H}_B$.)
Hence we obtain the following one-to-one
correspondence:
\begin{align*}
\psi\in{\cal H}_A\otimes{\cal H}_B\quad
\Leftrightarrow\quad
X\in B({\cal H}_B,{\cal H}_A),
\quad s.t.\quad TrX^*X<\infty,
\end{align*}
through the relation
\begin{align}
\psi
=\sum_{i}(Xf_i)\otimes f_i.
\label{asso}
\end{align}
Here $B({\cal H}_B,{\cal H}_A)$ indicates the set of bounded
operators from ${\cal H}_B$ to ${\cal H}_A$.

Now let us consider a set of orthonormal $M$ vectors 
$\psi_1,\cdots, \psi_M$ in ${\cal H}_A\otimes{\cal H}_B$.
We can associate each 
$\psi_l$ with an operator $X_l$ through (\ref{asso}).
As in (\ref{trace}), $X_m^*X_l$ 
are trace class operators on ${\cal H}_B$
for all $1\le m,l\le M$ and satisfy
\begin{align}
Tr X_m^*X_l
=\left\langle\psi_m,\psi_l\right\rangle=\delta_{m,l},\quad
1\le m,l\le M.
\label{otg}
\end{align}
Let ${\cal K}$ be the real linear subspace of trace class 
self-adjoint operators
on ${\cal H}_B$ spanned by operators
$\{X_m^*X_l+X_l^*X_m,\; i(X_m^*X_l-X_l^*X_m)\}_{m\neq l}$.
Let $N$ be the dimension of ${\cal K}$ 
and $(A_1,\cdots,A_{N})$
an arbitrary basis of ${\cal K}$.
The dimension $N$ is bounded as $N\le M(M-1)$.
Because each $X^*_mX_l$ satisfies (\ref{otg}), we have 
\begin{align}
Tr A_i=0,\quad i=1,\cdots,N.
\label{atrace}
\end{align}
We will call 
$\cal K$
the real vector space of trace class self-adjoint operators associated
with $\psi_1,\cdots,\psi_M$.

Now we are ready to state our main results.
In this paper, we show the following theorems:
\begin{thm}
Let ${\cal H}_A$, ${\cal H}_B$ 
be (possibly infinite dimensional) separable Hilbert spaces.
Let $\psi_1,\cdots,\psi_M$ be a set of orthogonal pure
states in ${\cal H}_A\otimes{\cal H}_B$ and 
$\cal K$ the associated real vector space of 
trace class self-adjoint operators
on ${\cal H}_B$.
Then if the dimension of ${\cal K}$ is $2$, 
the states $\psi_1,\cdots,\psi_M$ are distinguishable by LOCC
with certainty.
In particular, any pair of orthogonal pure
states $\psi_1,\psi_2$ 
are distinguishable by LOCC with certainty.
\label{fst}
\end{thm}
\begin{thm}
Let ${\cal H}_A$, ${\cal H}_B$ be (possibly infinite dimensional) separable Hilbert spaces.
Let $\psi_1,\cdots,\psi_M$ be a set of orthogonal pure
states in ${\cal H}_A\otimes{\cal H}_B$ and 
$\cal K$  the associated real vector space of trace class 
self-adjoint operators
on ${\cal H}_B$.
Suppose that the dimension of $\cal K$ is $3$.
Then
$\psi_1,\cdots,\psi_M$ are distinguishable by 
conclusive LOCC protocol with
probability $P_d$ such that
\[
P_d\ge 1-\underset{1\le l\le M}{max}\left(\sum_{k=1}^2p_k^l\right).
\]
Here, $p^l_k$ represents the $k$-th Schmidt 
coefficient of $\psi_l$,
ordered in the decreasing order.
\label{snd}
\end{thm}
\begin{rem}
The last statement of
Theorem \ref{fst} is the extension of \cite{WH} to infinite dimensional system.
Applying the argument in \cite{WH}, we can extend the result 
to multipartite systems:
any two orthogonal pure states in multipartite systems
are distinguishable by LOCC even in infinite dimensional systems. 
\end{rem}
\begin{rem}
In \cite{virmani}, S.Virmani et.al. showed that any two
(even non-orthogonal) 
multipartite pure states in finite dimensional systems
 can be optimally distinguished
using only LOCC.
It was derived using the result of the orthogonal case in
\cite{WH}.
The argument there can be applied to our infinite dimensional case.
Therefore, any two 
bipartite pure states
 can be optimally distinguished
using only LOCC, even for infinite dimensional system.
\end{rem}
\section{Proof}
\label{proof}
In this section, we prove the main theorems.
We correlate the problem of the distinguishability
with that of the convexity of the joint numerical ranges.
Let $(A_1,\cdots, A_N)$ be bounded self-adjoint operators 
on a Hilbert space $\cal H$.
A subset of ${\mathbb R}^N$ given by 
\begin{align*}
\left\{\left(
\langle z,A_1 z\rangle,\;\langle z,A_2 z\rangle,\;\cdots,
\langle z,A_N z\rangle
\right);\; z\in{\cal H},\;\Vert z\Vert=1\right\}
\subset{\mathbb R}^N
\end{align*}
is called the joint numerical range of $(A_1,\cdots,A_N)$.
Furthermore, for an orthogonal projection $P$ on $\cal H$,
we will call the set 
\begin{align*}
C_P(A_1,\cdots,A_N)\equiv
\left\{\left(
\langle z,A_1 z\rangle,\;\langle z,A_2 z\rangle,\;\cdots,
\langle z,A_N z\rangle
\right);\; z\in P{\cal H},\;\Vert z\Vert=1\right\}
\subset{\mathbb R}^N,
\end{align*}
the joint numerical range of $(A_1,\cdots,A_N)$ restricted
to the sub-Hilbert space $P{\cal H}$.
Theorem \ref{fst}, Theorem \ref{snd}
are derived as corollaries of the following propositions:
\begin{prop}
Let $\psi_1,\cdots,\psi_M$ 
be a set of orthogonal pure states in ${\cal H}_A\otimes {\cal H}_B$,
and $\cal K$ the associated real vector space of trace class self-adjoint
operators on ${\cal H}_B$.
Let $(A_1,\cdots, A_{N})$ be a basis of $\cal K$.
Suppose that for any projection $P$ on ${\cal H}_B$,
$C_P(A_1,\cdots,A_N)$ is convex.
Then the states $\psi_1,\cdots,\psi_M$ are distinguishable
by LOCC with certainty.
\label{thm1}
\end{prop}
\begin{prop}
Let $\psi_1,\cdots,\psi_M$ 
be a set of orthonormal pure states in ${\cal H}_A\otimes {\cal H}_B$,
and $\cal K$ the associated real vector space of trace class self-adjoint
operators on ${\cal H}_B$.
Let $(A_1,\cdots, A_{N})$ be a basis of ${\cal K}$.
Suppose that for any projection $P$ of ${\cal H}_B$
with dimension larger than $N_p$,
$C_P(A_1,\cdots,A_N)$ is convex.
Then the states $\psi_1,\cdots,\psi_M$ are distinguishable
by LOCC with the probability $P_d$ such that
\[
P_d\ge1-\underset{1\le l\le M}{\max}
\left(\sum_{k=1}^{N_p}p^l_k\right).\]
Here, $p^l_k$ represents the $k$-th Schmidt coefficient of $\psi_l$,
ordered in the decreasing order.
\label{thm2}
\end{prop}

First we prove the Proposition \ref{thm1}.
The proof consists of four steps:
{\it Step 1}. First, we show that if ${\cal H}_B$ has an orthonormal basis $\{g_k\}$
such that 
$\langle g_k,A_i g_k\rangle=0$ for all $i=1,\cdots,N$ and $k$,
then, $\psi_1,\cdots,\psi_N$ are distinguishable by LOCC
(Lemma \ref{base}).
{\it Step 2}. Second, using convex analysis, we show that if the joint numerical range
of $(A_1,\cdots,A_N)$ is convex, there exists 
at least one vector $z\in{\cal H}_B$
such that 
$\langle z,A_i z\rangle=0$ 
for all $i=1,\cdots,N$
(Lemma \ref{one}).
{\it Step 3}. Third, using Lemma \ref{one},
we show the existence of the orthonormal basis
satisfying the desired condition in {\it Step 1}
(Lemma \ref{convex}).
{\it Step 4}. Finally,
combining the results of {\it Step 1} and {\it Step 3},
we obtain Proposition \ref{thm1}.

Now let us start the proof.
First we show the following Lemma:
\begin{lem}
Let $(A_1,\cdots,A_N)$ be a basis of $\cal K$
associated with $\psi_1,\cdots,\psi_M$.
Suppose that there exists an orthonormal basis 
$\{g_k\}$ of ${\cal H}_B$
such that 
\begin{align}
\langle g_k,A_i g_k\rangle=0,\quad \forall k,\quad i=1\cdots N.
\label{asA}
\end{align}
Then the states $\psi_1,\cdots,\psi_M$ are distinguishable by LOCC.
\label{base}
\end{lem}
{\it Proof}   Let $\{f_i\}$ be the orthonormal basis 
fixed in Section \ref{rep}.
(Recall that we defined the operators $X_l$s in terms of $\{f_i\}$.)
We define an antilinear operator
$J:{\cal H}_B\to{\cal H}_B$ to be the complex conjugation with respect to
$\{f_i\}$: 
\begin{align*}
J\sum_i\alpha_i f_i\equiv\sum_i\bar{\alpha_i}f_i.
\end{align*}
As $J$ is an antilinear isometry, 
$\{Jg_k\}$ is an orthonormal basis of
${\cal H}_B$.
Therefore, we can decompose $\psi_1,\cdots,\psi_M$ with respect to
$\{Jg_k\}$:
\begin{align}
\psi_l=\sum_k\xi^l_k\otimes Jg_k.
\label{dcom1}
\end{align}
We show that for each $k$, $\{\xi_k^1,\cdots,\xi_k^M\}$ 
are mutually orthogonal.

Let us decompose $\psi_l$ with respect to $\{f_i\}$:
\begin{align}
\psi_l=\sum_i\varphi_i^l\otimes f_i.
\label{decom2}
\end{align}
Comparing (\ref{dcom1}) and (\ref{decom2}),
we obtain
\begin{align*}
\xi_k^l=\sum_i\varphi_i^l\langle Jg_k,f_i\rangle
=\sum_i\varphi_i^l\langle f_i, g_k\rangle
=X_lg_k.
\end{align*}
As $(A_,\cdots,A_N)$ is a basis of ${\cal K}$,
the assumption (\ref{asA}) 
implies
\begin{align*}
\langle \xi^l_k, \xi_k^{m}\rangle=
\langle X_l g_k, X_m g_k\rangle=0
\quad \forall\;l\neq m,\quad \forall k.
\end{align*}
Hence for each $k$, $\{\xi_k^1,\cdots,\xi_k^M\}$ 
are mutually orthogonal.

Thus (\ref{dcom1}) takes the form of (\ref{decom}), with 
the orthogonality condition (\ref{ortho}).
Therefore, from the arguments in the Introduction,
we can distinguish $\psi_1,\cdots,\psi_M$ by LOCC
with certainty.
$\square$\\\\
Next we show the following Lemma
which holds on a general Hilbert space ${\cal H}$:
\begin{lem}
Let $(A_1,\cdots,A_N)$ be a set of trace class self-adjoint 
operators on a Hilbert space $\cal H$ such that
$Tr A_i=0$ for each $1\le i\le N$.
Suppose that the joint numerical range of 
$(A_1,\cdots,A_N)$ is a convex subset of ${\mathbb R}^N$.
Then there exists a vector $z\in{\cal H}$ with $\Vert z \Vert=1$
such that
\begin{align*}
\langle z, A_i z\rangle=0,\quad i=1,\cdots, N.
\end{align*}
\label{one}
\end{lem}
{\it Proof}\\
Before starting the proof, we review some basic facts from 
convex analysis \cite{fund}.
Let $x_1,\cdots, x_k$ be elements in ${\mathbb R}^N$.
An element 
$\sum_{i=1}^{k}\alpha_i x_i$ with real coefficients $\alpha_i$
satisfying $\sum_{i=1}^{k}\alpha_i=1$ 
is called an affine combination of $x_1,\cdots,x_k$.
An affine manifold in
${\mathbb R}^N$ is a set containing all its affine combinations.
Let $S$ be a nonempty subset of ${\mathbb R}^N$.
The affine hull of $S$ is defined to be 
the smallest affine manifold
containing $S$.
We denote the affine hull of S by ${\rm aff} S$.
In other words, ${\rm aff} S$ is the affine 
manifold generated by $S$.
As easily seen, it is a closed plane
parallel to a linear subspace 
in ${\mathbb R}^N$.
Its dimension may be lower than $N$ in general.
The relative interior of $S$, ${\rm ri}S$,
is the interior of $S$ with respect to the topology
relative to ${\rm aff}S$.
In other words, 
\begin{align*}
{\rm ri} S\equiv\{x\in S;\;\exists\; \varepsilon >0\;
s.t.\;
B(x,\varepsilon )\cap{\rm aff}S\subset S\}.
\end{align*}
Here, $B(x,\varepsilon)$ is a ball of radius $\varepsilon$, 
centered at $x$.
The following fact is known:
\begin{lem}
Let $C$ be a nonempty convex subset of ${\mathbb R}^N$.
Then for any point $x_0$ in ${\rm aff}C\backslash{\rm ri}C$,
there exists a non-zero vector $s\in {\mathbb R}^N$ 
parallel to ${\rm aff}C$, such that 
\[
\left\langle\left\langle s,x-x_0\right\rangle\right\rangle
\ge 0,\quad \forall x\in C.
\]
Here $\left\langle\left\langle\;,\;\right\rangle\right\rangle$
is the inner product of ${\mathbb R}^N$:
\begin{align*}
\left\langle\left\langle s, x\right\rangle\right\rangle
\equiv\sum_{i=1}^Ns_i\cdot x_i.
\end{align*}
\label{hyper}
\end{lem}
Now we are ready to prove Lemma \ref{one}.
The claim is equivalent to saying
that $0$ is included in
the joint numerical range
of the operators $(A_1,\cdots, A_N)$.
We denote the joint numerical range by $C_1$:
\begin{align*}
C_1\equiv\left\{\left(
\langle z,A_1 z\rangle,\;\langle z,A_2 z\rangle,\;\cdots,
\langle z,A_N z\rangle
\right)\in{\mathbb R}^N;\; z\in{\cal H},\;\Vert z\Vert=1\right\}.
\end{align*}
By assumption, $C_1$ is a nonempty convex subset of ${\mathbb R}^N$.
Let $\{e_k\}$ be an arbitrary orthonormal basis of 
$\cal H$. By the definition of $C_1$,
\begin{align*}
x_k\equiv\left(\left\langle e_k,A_1 e_k\right\rangle,
\cdots,\left\langle e_k,A_N e_k\right\rangle\right)
\end{align*} 
is an element of $C_1$
for each $k$.

The finite dimensional case ${\cal H}={\mathbb C}^n$ is immediate.
By the convexity of $C_1$, we obtain
\begin{align*}
0=\frac{1}{n}\left(
Tr A_1,\cdots,Tr A_N
\right)
=\frac{1}{n}\sum_{k=1}^n
\left(
\langle e_k, A_1 e_k\rangle,
\cdots,
\langle e_k, A_N e_k\rangle
\right)
\in 
C_1.
\end{align*}
Below we prove the infinite dimensional case.

First we observe that $0$ is included in the closure of
$C_1$.
In particular, $0$ is in ${\rm aff}C_1$.
To see this, note that for all $l\in{\mathbb N}$, 
we have
\begin{align*}
\frac{1}{l}\sum_{k=1}^l
\left(
\langle e_k, A_1 e_k\rangle,
\cdots,
\langle e_k, A_N e_k\rangle
\right)\in C_1.
\end{align*}
As $A_i$ is a trace class operator,
the sum 
$\sum_{k=1}^\infty\langle e_k, A_i e_k\rangle$
converges absolutely.
By taking $l\to\infty$ limit, we obtain
\begin{align*}
0=\lim_{l\to\infty}\frac{1}{l}\sum_{k=1}^l
\left(
\langle e_k, A_1 e_k\rangle,
\cdots,
\langle e_k, A_N e_k\rangle
\right)\in 
\overline{C_1}\subset{\rm aff}C_1.
\end{align*}
Hence $0$ is in ${\rm aff}C_1$. 

Second, we show that $0$ is actually in ${\rm ri}C_1$. 
To prove this, assume $0$ is not included in ${\rm ri}C_1$.
Then it is an element of 
${\rm aff}C_1\backslash{\rm ri} C_1$.
As $C_1$ is a nonempty convex set,
from Lemma \ref{hyper},
there exists a non-zero vector 
$s=(s_1,\cdots,s_N)\in {\mathbb R}^N$ pararell
to ${\rm aff}C_1$, such that 
\begin{align*}
\left\langle\left\langle s,x\right\rangle\right\rangle
\ge 0,\quad \forall x\in C_1.
\end{align*}
As $x_k\in C_1$, we have
\begin{align}
\left\langle\left\langle s, x_k
\right\rangle\right\rangle
\ge 0,
\label{each}
\end{align}
for all $k$.
On the other hand, we have
\begin{align}
\sum_{k=1}^{\infty} \left\langle\left\langle s, x_k
\right\rangle\right\rangle
=\sum_{i=1}^N
s_i\sum_{k=1}^\infty \cdot \left\langle e_k,A_i e_k\right\rangle
=\sum_{i=1}^N s_i\cdot TrA_i=0.
\label{sum}
\end{align}
From (\ref{each}) and (\ref{sum}), we obtain
$\left\langle\left\langle s, x_k\right\rangle\right\rangle=0$ for all $k$.
As the orthonormal basis
$\{e_k\}$ can be taken arbitrary, 
we obtain
\begin{align*}
\left\langle\left\langle s, x
\right\rangle\right\rangle=0,\quad
\forall x\in C_1.
\end{align*}
As $s$ is a non-zero vector parallel to ${\rm aff}C_1$, 
this means that $C_1$ is included in some affine manifold
that is strictly smaller than ${\rm aff}C_1$.
This contradicts the definition of ${\rm aff}C_1$.
(Recall that ${\rm aff}C_1$ is 
the smallest affine manifold including $C_1$.)
Therefore, we obtain $0\in {\rm ri}C_1$.
In particular, $0\in C_1$ and this completes the proof.
$\square$
\\\\
Using Lemma \ref{one}, we obtain the following lemma: 
\begin{lem}
Let $(A_1,\cdots ,A_N)$ be a set of
trace class self-adjoint operators on
a Hilbert space $\cal H$ such that
$Tr A_i=0$ for each $1\le i\le N$. 
Suppose that for every orthogonal projection $P$ on $\cal H$,
$C_P(A_1,\cdots,A_N)$ is convex.
Then there exists an orthonormal basis $\{g_k\}$
of $\cal H$, such that
\[
\left\langle g_k,A_i g_k \right\rangle=0,\quad
\forall i=1,\cdots N,\quad \forall k.
\]
\label{convex}
\end{lem}
{\it Proof}\\
We will say that a set of vectors $Z$ in $\cal H$ satisfies {\it Property 
\nolinebreak*}
if it satisfies the following conditions:\\\\
{\it Property *}
\begin{enumerate}
\item $Z$ is a set of mutually orthogonal 
unit vectors of $\cal H$.
\item 
$\left\langle z,A_i z \right\rangle=0,\quad
i=1,\cdots,N\;$  for all $z\in Z$.
\end{enumerate}
By Zorn's lemma, there exists a maximal set of 
orthonormal vectors $\{g_k\}$ in
$\cal H$ which satisfies the {\it Property *}.
It suffices to show that $\{g_k\}$ is complete.

Suppose that $\{g_k\}$ is not complete in $\cal H$,
and let $P$ be the orthogonal 
projection onto the sub-Hilbert space 
spanned by $\{g_k\}$.
From {\it Property *}, we have
\begin{align*}
Tr PA_iP=\sum_k \left\langle g_k, A_i g_k\right\rangle=0,
\quad i=1,\cdots N.
\end{align*}
Let $\bar P$ be $\bar P=1-P$.
Now we regard $(\bar P A_1 \bar P,\cdots,\bar P A_N \bar P)$ as
self-adjoint trace class operators on the Hilbert space
${\bar P}{\cal H}$ such that
\begin{align*}
Tr_{\bar P{\cal H}}(\bar P A_i \bar P)
=Tr(A_i)-Tr(PA_iP)=0,\quad i=1,\cdots N.
\end{align*}
By the assumption,
the joint numerical range of 
$(\bar P A_1 \bar P,\cdots,\bar P A_N \bar P)$ 
on $\bar{P}{\cal H}$ is convex.
Thus, applying Lemma \ref{one},
there exists a unit vector $z\in\bar{P}{\cal H}$ such that
$\left\langle z, A_i z\right\rangle=0$ for all $i=1,\cdots,N$.
As $z$ is orthogonal to all $g_k$,
the set $\{z\}\cup\{g_k\}$ 
satisfies the {\it Property *}, and is strictly larger than
$\{g_k\}$.
This contradicts the maximality of $\{g_k\}$.
Therefore, $\{g_k\}$ is complete.
$\square$
\\\\
Now, let us complete the proof of Proposition \ref{thm1}.
The basis of $\cal K$, $(A_1,\cdots, A_{N})$
are trace class self-adjoint operators satisfying
$TrA_i=0,\;i=1,\cdots,N$ (\ref{atrace}).
Therefore, if $C_P(A_1,\cdots,A_N)$ is a convex subset
of ${\mathbb R}^N$ for any orthogonal projection 
$P$ on ${\cal H}_B$, 
there exists an orthonormal 
basis $\{g_k\}$ of ${\cal H}_B$ such that
$\left\langle g_k,A_i g_k \right\rangle=0$,
for all 
$i=1,\cdots N$ and $k$,
from Lemma \ref{convex}.
By Lemma \ref{base}, this concludes that $\psi_1,\cdots,\psi_M$
are distinguishable by LOCC.$\square$
\\\\
Proposition \ref{thm2} can be shown in the same way.
We have the following lemma:
\begin{lem}
Let $(A_1,\cdots ,A_N)$ be a set of trace class self-adjoint operators on
a Hilbert space $\cal H$ such that
$Tr A_i=0$ for each $1\le i\le N$. 
Suppose that for every orthogonal
projection $P$ on $\cal H$ with dimension
larger than $N_p$, $C_P(A_1,\cdots,A_N)$
is convex.
Then there exists an orthonormal basis 
$\{g_k\}$ of $\cal H$, such that
\[
\left\langle g_k,A_i g_k \right\rangle=0,\quad
i=1,\cdots N,\quad\forall k>N_p.
\]
\end{lem}
{\it Proof}\\
The same as the proof of Lem 3.6.
We can find a set of orthonormal vectors
satisfying {\it Property *},
such that the dimension of its complementary subspace is $N_p$.
$\square$\\
Decomposing each $\psi_l$ with respect to 
$\{Jg_k\}$,
we obtain
\begin{align}
\psi_l=\sum_{k}\xi^l_k\otimes J g_k.
\label{er}
\end{align}
By the argument in the proof of Lemma \ref{base},
(\ref{er}) takes the form of (\ref{decom}) 
with the orthogonality condition (\ref{prob}).
Therefore,
for the protocol in the Introduction,
the probability that the error occurs is
$\sum_{k=1}^{N_p}\Vert\xi^l_k\Vert^2$
when $\psi=\psi_l$.
It is bounded from above as follows:
\begin{align*}
&\sum_{k=1}^{N_p}\Vert\xi^l_k\Vert^2
=\sum_{k=1}^{N_p}
\left\Vert\left( 1\otimes\ket{Jg_k}\bra{Jg_k}\right)
\psi_l\right\Vert^2
\\
&\le
\sup\left\{\sum_{k=1}^{N_p}
\left\Vert\left( 1\otimes\ket{z_k}\bra{z_k}\right)\psi_l\right\Vert^2;
\;\left\{z_k\right\}_{k=1}^{N_p} :\;\;{\rm orthonormal\; set\; of\;}
{\cal H}_B
\right\}
=
\sum_{k=1}^{N_p} p_k^{l}.
\end{align*}
Here, $p^l_k$ is the $k$-th Schmidt coefficient of $\psi_l$,
ordered in the decreasing order.
Therefore,
$\psi_1,\cdots,\psi_M$ are distinguishable by LOCC with
probability $P_d$ such that
\[
P_d
\ge 1-\underset{1\le l\le M}{\max}
\left(
\sum_{k=1}^{N_p} \Vert\xi_k^l\Vert^2
\right)
\ge
1-\underset{1\le l\le M}{\max}
\left(\sum_{k=1}^{N_p}p_k^l\right),
\]
and we obtain Proposition \ref{thm2}.$\square$
\\\\
{\it Proof of Theorem \ref{fst} and Theorem \ref{snd}}
\\\\
Now we apply the known results about joint numerical range to
Proposition \ref{thm1}, \ref{thm2}
and derive Theorem \ref{fst} and Theorem \ref{snd}.
For $N=2$ case,
the following Theorem is known \cite{Hal}:
\begin{thm}
For any bounded self-adjoint operators $T_1,T_2$ 
on a separable Hilbert space $\cal H$,
the set
\begin{align*}
\left\{
\left(
\left\langle z,T_1 z\right\rangle, 
\left\langle z,T_2 z\right\rangle\right)
\in{\mathbb R}^2,\quad
z\in {\cal H},\quad
\Vert z\Vert=1
\right\}
\end{align*}
is a convex subset of ${\mathbb R}^2$.
\label{top}
\end{thm}
This is called Toeplitz Hausdorff Theorem.
By this Theorem, $C_P(A_1,A_2)$ is a convex subset
of ${\mathbb R}^2$ for any projection $P$ on ${\cal H}_B$.
Therefore, 
applying Proposition \ref{thm1}, we obtain Theorem \ref{fst}.
The last statement comes from the fact $N\le M(M-1)=2$ for $M=2$.

On the other hand, for $N=3$, the next Theorem is known \cite{bind},\cite{fan}.
\begin{thm}
Let $\cal H$ be a separable
Hilbert space with $dim {\cal H}\ge 3$.
Then for any self-adjoint operators $T_1,T_2,T_3$
in $\cal H$,
the set
\begin{align*}
\left\{
\left(
\left\langle z,T_1 z\right\rangle, 
\left\langle z,T_2 z\right\rangle,
\left\langle z,T_3 z\right\rangle
\right)
\in{\mathbb R}^3,\quad
z\in {\cal H},\quad
\Vert z\Vert=1
\right\}
\end{align*}
is a convex subset of ${\mathbb R}^3$.
\label{three}
\end{thm}
By this Theorem, $C_P(A_1,A_2,A_3)$ is a convex subset
of ${\mathbb R}^3$ for any projection $P$ on ${\cal H}_B$
with dimension larger than $2$.
Therefore, 
applying Proposition \ref{thm2}, we obtain Theorem \ref{snd}.
$\square$.
\\\\
\noindent
{\bf Acknowledgement.}\\
{
This work is supported by
Research Fellowships of the Japan Society for 
the Promotion of Science for Young
Scientists.
}

\end{document}